\documentclass[pre,eqsecnum,preprint,showpacs]{revtex4}
\usepackage{amsmath}
\usepackage{epsfig}
\usepackage{graphicx}

\def\hangin{\begin{list}{}{\setlength{\leftmargin}{1.1in}
\setlength{\itemindent}{-1.1in}}}
\def\endhangin{\end{list}}

\begin{document}

\noindent{\Large GRANULAR FLUIDS}

\vspace{0.5in}

\noindent James W. Dufty

\noindent Department of Physics, University of Florida, Gainesville, FL 32611

\vspace{0.3in}

\noindent\textbf{ARTICLE OUTLINE}

Definition of the Subject and Importance

Introduction

Granular Matter and its Statistical Mechanics

\qquad Nonequilibrium statistical mechanics

\qquad Liouville equation

\qquad Stationary homogeneous state

Macroscopic Balance Equations

"Normal" States and Hydrodynamics

Navier-Stokes Approximation

\qquad Constitutive equations

\qquad Green-Kubo expressions

\qquad Hydrodynamic equations

Discussion and Outlook

Bibliography

Appendix - Gradient Expansion

\vspace{0.3in}

\section{ Definition of the Subject and its Importance}

\label{sec1}

The terminology granular matter refers to systems with a large number of
hard objects (grains) of mesoscopic size ranging from millimeters to meters.
Geological examples include desert sand and the rocks of a landslide. But
the scope of such systems is much broader, including powders and snow,
edible products such a seeds and salt, medical products like pills, and
extraterrestrial systems such as the surface regolith of Mars and the rings
of Saturn. The importance of a fundamental understanding for granular matter
properties can hardly be overestimated. Practical issues of current concern
range from disaster mitigation of avalanches and explosions of grain silos
to immense economic consequences within the pharmaceutical industry. In
addition, they are of academic and conceptual importance as well as examples
of systems far from equilibrium.

Under many conditions of interest, granular matter flows like a normal fluid
\cite{Kadanoff}. In the latter case such flows are accurately described by
the equations of hydrodynamics. Attention is focused here on the possibility
for a corresponding hydrodynamic description of granular flows. The tools of
nonequilibrium statistical mechanics \cite{McL}, developed over the past
fifty years for fluids composed of atoms and molecules \cite{Hansen,Resibois}%
, are applied here to a system of grains for a fundamental approach to both
qualitative questions and practical quantitative predictions. Applications
of basic atomic physics principles to granular fluids have accelerated
during the past decade, starting with an emphasis on molecular dynamics (MD)
simulations \cite{HCSMD} and kinetic theory \cite{Poschel2,Dufty(KT)01}, and
more recently with the theoretical methods of the type described here \cite%
{Brey97,vanN01,DBL02,Dufty00,DBB06,DBB07}.

\section{Introduction}

\label{sec2}

To start with the familiar, consider a jar of vitamin pills, mustard seeds,
or peanuts. Remove the lid and pour them into a bowl, observing that the
"flow", or their collective motion, has some similarity to that of a normal
fluid such as water. The collective motion in both cases is the consequence
of collisions among their constituents, grains or atoms, and their large
number. It is tempting to make the correspondence of grains to atoms in
considering the similarities of flows in these two types of fluids. The
objective here is to explore in formal detail the extent to which that
correspondence is conceptually and quantitatively justified. An important
prerequisite is the integrity of the grains during their motion. Each grain
is comprised of a large number of atoms or molecules. Integrity refers to
their retention of mass and shape following interactions with other grains
or with their environment. As such, the grains behave as \textquotedblleft
particles\textquotedblright\ whose detailed internal structure is not
essential to their description, which is captured instead by a few
parameters describing their shape, mass, and collisional properties with
other grains. However, an important consequence of this underlying molecular
structure is a redistribution of translational kinetic energy of the grains
and internal energy of the constituent molecules. At the mesoscopic level
this appears as an energy loss on collisions between pairs of grains. This
is a central feature of granular fluids differentiating them from atomic
fluids: the inelasticity of granular pair collisions.

Granular matter occurs in two classes of states, compact and activated \cite%
{Halsey}. In the first case, the grains form a static packed configuration
within the container due to the effects of gravity on their relatively large
mass and their inelastic collisions. Any initial motion is quickly
dissipated and their kinetic energy becomes negligible relative to the
gravitational potential energy. Important questions arise about the possible
and probable packing configurations that determine the stresses within the
system and the distribution of forces on the container. For example, chains
of particles in contact can occur as arches to support matter above them
while reducing their force on the matter below. There is an intense interest
in the study of such states, generically referred to as contact mechanics.

Activated states refer to continuously driven systems, or gravity free
conditions. For example, a container of grains in a compact configuration
can be shaken to impose kinetic energy and motion among the grains.
Similarly, unrestrained systems in a gravitational field will flow towards
lower potential energy (e.g., hopper flow or flow down an incline). Initial
activation in space laboratory experiments provides another example
(self-sustained fluidization). For the flows considered here as candidates
for a hydrodynamic description continual collisions are essential. This
means that within each small cell, still containing many particles, the
particles are moving randomly relative to the collective motion of that
cell. Thus, ballistic motion or beams with all particles moving
independently in the same direction are excluded.

Both compact and activated grains may occur immersed in a continuum such as
water or air that may have a strong or weak effect on their collective
properties. For compact systems water may provide a lubrication effect that
affects the dominant class of configurations. For activated systems it can
provide an additional dissipative drag between collisions among the grains.
When the medium plays an important role the systems is said to be wet. In
the opposite limit it is said to be dry. Finally, it is possible for compact
and activated components of a system to coexist as heterogeneous states.
Here, only the simplest case of dry systems in fully activated flows are
considered. These are referred to in the following as granular fluids.

Advances in the study of granular fluids have arisen from many communities,
including chemical engineering, materials sciences, and physics. The
additional academic and conceptual importance of granular matter as
practical systems for exploring the relevance of many-body fluid methods is
primarily for the physics community. Granular matter, viewed as a system of
particles with inelastic interactions, provides new opportunities to test
the qualitative and quantitative limits of many-body methods developed over
the past century for atomic and molecular systems. This is the field of
non-equilibrium statistical mechanics \cite{McL,Hansen,Resibois}. Granular
matter provides a new testing ground for a reconsideration of the most
fundamental concepts and tools \cite{Kadanoff}, with the potential for
enhanced understanding of their place in atomic and molecular systems as
well.

Statistical mechanics addresses the difficult many-body problem of
extracting macroscopic properties of experimental interest from the very
large number of constituent particles. The results express these properties
in terms of the fundamental "microscopic" features of these particles, such
as mass, shape, degree of inelasticity, and collisional properties. In this
way the properties of the vitamin pills, mustard seeds, and peanuts are
distinguished at a fundamental level. Also, conceptual issues such as the
limitations of a macroscopic description are exposed through the insistence
on their logical evolution from the fundamental microdynamics. In the next
section, the granular fluid is described as a system of particles
interacting via pairwise additive, nonconservative forces. The microscopic
dynamics of these particles leads to balance equations for the mass density,
energy density, and momentum density. Their averages define the
"hydrodynamic fields" which are candidates for a macroscopic, continuum
mechanics description. These exact equations are described in Section 3 and
the need for "constitutive equations" to provide a closure is described. The
origin of constitutive equations, and consequently the origin of
hydrodynamics, is associated with the concept of "normal states" \cite%
{Origins} in Section 4. The normal state for the case of small spatial
deviations from homogeneity is constructed formally in Section 5, resulting
in the constitutive equations for Navier-Stokes hydrodynamics \cite%
{Navier-Stokes}. This derivation also provides insight into the context in
which such a description should hold, and differences from the corresponding
results for a normal fluid are noted. Empirical evidence \cite%
{Caldera,Swinney}, simulations \cite{BRMC99,Brey,Brey(GK)}, and
corresponding results from kinetic theory \cite{BDKyS98,DB(GK)01,Bari,Modes}
support the applicability of this hydrodynamic description under appropriate
conditions. Finally, the results are summarized in Section 6 and some
comments on the outlook for future developments are offered.

The presentation here is focused on recent work of the author and his
collaborators for application of statistical mechanics to explore
hydrodynamics for a granular gas. Consequently, the references quoted are
heavily weighted toward those developmental studies. Apologies are offered
for the exclusion of the vast and important complementary literature on
simulations, kinetic theory, and experiments also bearing on this topic.
Many of these can be found in the list of Books and Reviews given here.

\section{Granular Fluid and its Statistical Mechanics}

\label{sec3}

\subsection{Nonequilibrium statistical mechanics}

Consider a system of $N>>1$ identical grains (hereafter referred to as
particles) in a volume $V$, whose initial positions $\{\mathbf{q}_{i}\}$ and
velocities $\{\mathbf{v}_{i}\},$ $1\leq i\leq N,$ are specified. The
positions and velocities define a point in a $6N$ dimensional space denoted
by $\Gamma \equiv \{\mathbf{q}_{i},\mathbf{v}_{i}\}$, defining the
microstate of the system. A macrostate is defined by a probability density $%
\rho \left( \Gamma \right) $ in this space, representing statistical rather
than precise knowledge of the system. The field of statistical mechanics
addresses properties of macrostates, based on the recognition that for very
large $N$ the details of microstates are neither experimentally accessible
nor practically calculable. Properties of interest are represented by
functions $A(\Gamma )$, and their values for a macrostate $\rho \left(
\Gamma \right) $ are determined from the expectations%
\begin{equation}
\left\langle A;\rho \right\rangle \equiv \int d\Gamma \rho \left( \Gamma
\right) A\left( \Gamma \right) .  \label{3.1}
\end{equation}%
In this section, a brief overview of the essential ingredients of
nonequilibrium statistical mechanics is given, broadened from its usual form
\cite{McL} to include granular matter.

The dynamics of macrostates is determined from the underlying dynamics of
the microstates. The initial point $\Gamma $ changes in time since the
particles have velocities and move to new positions. They move in straight
lines until one or more come within the force field of other particles, at
which point their velocities change as well as their positions. The forces
are taken to be pairwise additive, such that the total force on particle $i$
is $\mathbf{F}_{i}=\sum_{j}\mathbf{F}_{ij}$, where $\mathbf{F}_{ij}$ is the
force on particle $i$ due to particle $j$. This does not mean that the
interactions are pairwise sequential; three or more particles can interact
simultaneously. The pair forces are restricted by Newton's third law, $%
\mathbf{F}_{ij}=-\mathbf{F}_{ji}$, with conservation of momentum. Otherwise
quite general forces can be considered to represent the shape of the
particles and their degree of inelasticity. It is assumed here that the
force range vanishes outside a distance $\sigma /2$ from the center of each
particle so that $\sigma $ characterizes the size of the particles.
Furthermore, the particles are taken to be strongly repulsive so that their
mean maximum overlap $d$ on collision is small compared to their size, $%
d/\sigma <1$. However, their size can be large or small compared to the mean
distance between particles $\left( V/N\right) ^{1/3}$, depending on whether
the density of the system is small or large, respectively. Most importantly
for the purposes here, these forces do not conserve energy. This property
captures the feature of real grains that center of mass kinetic energy is
lost as they distort during pair collisions. Further details of the force
law are not required at this point.

The dynamics consists of straight line motion along the direction of the
velocity at time $t$ (free streaming), until the force range of any pair of
particles, say $i,j$, overlaps. The relative velocity $\mathbf{g}_{ij}=%
\mathbf{v}_{i}-\mathbf{v}_{j}$ of that pair changes according to Newton's
second law for the chosen force law $\mathbf{F}_{ij}.$ Subsequently, all
particles continue to stream freely until another pair has a force range of
overlap, and the collisional change is repeated for that pair. In this way a
trajectory $\Gamma _{t}\equiv \left\{ \mathbf{q}_{1}(t),\ldots ,\mathbf{q}%
_{N}(t),\mathbf{v}_{1}(t),\dots ,\mathbf{v}_{N}(t)\right\} $ is generated
for $t>0$. This trajectory is unique and invertible. The statistical
mechanics for a fluid of inelastic particles \cite%
{Brey97,Dufty00,vanN01,DBL02} is comprised of the dynamics just described, a
macrostate specified in terms of a probability density $\rho (\Gamma )$, and
a set of observables generically denoted by $A(\Gamma )$. The expectation
value for an observable at time $t>0$ for a state $\rho (\Gamma )$ given at $%
t=0$ is defined by
\begin{equation}
\langle A(t);0\rangle \equiv \int d\Gamma \rho (\Gamma )A(\Gamma _{t})\equiv
\int d\Gamma \,\rho (\Gamma )e^{tL}A(\Gamma )  \label{3.4}
\end{equation}%
where $A(t)=A(\Gamma _{t})$, and $\Gamma _{t}\equiv \left\{ \mathbf{q}%
_{1}(t),\ldots ,\mathbf{q}_{N}(t),\mathbf{v}_{1}(t),\dots ,\mathbf{v}%
_{N}(t)\right\} $ is the phase point evolved to time $t$ from $\Gamma
=\Gamma _{t=0}$. The dynamics can be represented in terms of a generator $L$
defined by the second equality of (\ref{3.4}). There are two components to
the generator, corresponding to the two steps of free streaming and velocity
changes due to interactions
\begin{equation}
L=\sum_{i=1}^{N}\mathbf{v}_{i}\cdot \mathbf{\nabla }_{i}+\frac{1}{2m}%
\sum_{i=1}^{N}\sum_{j\neq i}^{N}\mathbf{F}_{ij}\cdot \left( \boldsymbol{%
\nabla }_{\mathbf{v}_{i}}-\boldsymbol{\nabla }_{\mathbf{v}_{j}}\right) .
\label{3.6}
\end{equation}

An alternative equivalent representation of the dynamics is obtained by
transferring the dynamics from the observable $A(\Gamma )$ to the state $%
\rho (\Gamma )$ by the definition
\begin{equation}
\int d\Gamma \,\rho (\Gamma )e^{tL}A(\Gamma )\equiv \int d\Gamma \,\left(
e^{-t\overline{L}}\rho (\Gamma )\right) A(\Gamma )\equiv \int d\Gamma \,\rho
(\Gamma ,t)A(\Gamma ).  \label{3.10}
\end{equation}%
The representation in terms of a dynamical state $\rho (\Gamma ,t)$ is
referred to as Liouville dynamics. Its generator $\overline{L}$ is the
formal adjoint of $L$ which is found to be%
\begin{equation}
\overline{L}=L+\frac{1}{2m}\sum_{i=1}^{N}\sum_{j\neq i}^{N}\left(
\boldsymbol{\nabla }_{\mathbf{v}_{i}}-\boldsymbol{\nabla }_{\mathbf{v}%
_{j}}\right) \cdot \mathbf{F}_{ij}  \label{3.11}
\end{equation}%
The difference between $L$ and $\overline{L}$ arises because the forces are
non-conservative and therefore depend on the relative velocities of each
pair as well as there positions. Time correlation functions for two
observables $A$ and $B$ are defined in a similar way%
\begin{equation}
\langle A(t)B;0\rangle \equiv \int d\Gamma \left( e^{tL}A(\Gamma )\right)
\rho (\Gamma )B(\Gamma )=\int d\Gamma A(\Gamma )\left( e^{-t\overline{L}%
}\rho (\Gamma )\right) \left( e^{-tL}B(\Gamma )\right) .  \label{3.13}
\end{equation}%
or
\begin{equation}
\langle A(t)B;0\rangle \equiv \langle AB(-t);t\rangle  \label{3.14}
\end{equation}

In summary, averages like $\langle A(t);0\rangle $ and correlation functions
$\langle A(t)B;0\rangle $ are the central properties of interest for a
macroscopic description of physical systems. The microscopic dynamics can be
represented in terms of the observables $A(\Gamma ,t)$ or the states $\rho
(\Gamma ,t)$ which are determined from specified initial values and the
equations
\begin{equation}
\left( \partial _{t}-L\right) A(\Gamma ,t)=0,\hspace{0.3in}\left( \partial
_{t}+\overline{L}\right) \rho (\Gamma ,t)=0.  \label{3.15}
\end{equation}%
In the following most of the analysis is done in terms of the of states, and
$\,$ the associated equation of motion is known as the Liouville equation.

\subsection{Liouville equation and cooling}

For an isolated system, the total energy decreases monotonically due to the
loss of energy on each pair collision. This is reflected in a decrease of
the average kinetic energy of the particles between collisions and hence is
referred to as collisional "cooling". The energy per particle at time $t$
and its loss are%
\begin{equation}
\epsilon \left( t\right) \equiv N^{-1}\langle E;t\rangle ,\hspace{0.3in}%
\omega (t)\equiv -\partial _{t}\epsilon \left( t\right) =N^{-1}\langle
LE;t\rangle .  \label{3.16}
\end{equation}%
This cooling effect is common to all solutions to the Liouville equation and
it is useful to separate the dynamics into that due to this cooling and the
residual time dependence%
\begin{equation}
\rho (\Gamma ,t)\equiv \rho (\Gamma ,\epsilon (t),t).  \label{3.17}
\end{equation}%
The Liouville equation then can be written%
\begin{equation}
\partial _{t}\rho (\Gamma ,\epsilon ,t)\mid _{\epsilon }+\left( -\omega
\left( \epsilon ,t\right) \partial _{\epsilon }+\overline{L}\right) \rho
(\Gamma ,\epsilon ,t)=0.  \label{3.18}
\end{equation}%
The time derivative is now taken at constant $\epsilon $. The notation $%
\omega \left( \epsilon ,t\right) $ reflects the fact that it is a linear
functional of $\rho (\Gamma ,\epsilon ,t)$, from its definition (\ref{3.16}%
). This is a useful form that isolates a primary effect of the
nonconservative forces (cooling) from the residual dynamics that will be
associated with relaxation of the spatial inhomogeneities of interest below.
For notational simplicity (\ref{3.18}) is written
\begin{equation}
\left( \partial _{t}+\overline{\mathcal{L}}\right) \rho (\Gamma ,\epsilon
,t)=0,\hspace{0.3in}\overline{\mathcal{L}}\equiv -\omega \left( \epsilon
,t\right) \partial _{\epsilon }+\overline{L}.  \label{3.19}
\end{equation}%
The corresponding equation for observables is%
\begin{equation}
\left( \partial _{t}-\mathcal{L}\right) A(\Gamma ,\epsilon ,t)=0,\hspace{%
0.3in}\mathcal{L}\equiv -\partial _{\epsilon }\omega \left( \epsilon
,t\right) +L.  \label{3.20}
\end{equation}%
where it is understood that $\partial _{\epsilon }$ operates on everything
to its right.

\subsection{Stationary homogeneous state}

An isolated normal fluid supports an equilibrium state. This is a stationary
solution to the Liouville equation with translational invariance, the Gibbs
states. From the discussion above it is clear that isolated granular fluids
have no truly stationary state due to cooling. However, there is a
"universal" homogeneous state similar to the Gibbs state in the sense that a
wide class of homogeneous initial states rapidly approach this state, on the
time scale of a few collisions per particle. It is simple in the sense that
all of its time dependence is that associated with cooling%
\begin{equation}
\rho _{0}(\Gamma ,t)=\rho _{0}(\{\mathbf{q}_{ij},\mathbf{v}_{i}\},\epsilon
(t)).  \label{3.21}
\end{equation}%
Here, $\mathbf{q}_{ij}=\mathbf{q}_{i}-\mathbf{q}_{j}$ so the solution also
has translational invariance. In the representation (\ref{3.19}) it is seen
to be a stationary solution to the Liouville equation
\begin{equation}
\overline{\mathcal{L}}\rho _{0}=0.  \label{3.37}
\end{equation}%
There is no longer any explicit time dependence since $\omega \left(
\epsilon ,t\right) =\omega \left( \epsilon \right) $ and $\overline{\mathcal{%
L}}=-\omega \left( \epsilon \right) \partial _{\epsilon }+\overline{L}$ for
this state. This solution is referred to as the homogeneous cooling state
(HCS). Clearly, it is the close analogue of the Gibbs state for a normal
fluid. It is an example of a "normal" state in the sense that all of its
time dependence occurs through one of the hydrodynamic fields (the energy).
This concept is sharpened below.

\section{Macroscopic Balance Equations}

\label{sec4}

The origins of a macroscopic description for a fluid are the balance
equations for the average mass density $\left\langle m\left( \mathbf{r}%
\right) ;t\right\rangle $, energy density $\left\langle e\left( \mathbf{r}%
\right) ;t\right\rangle $, and momentum density $\left\langle \mathbf{g}%
\left( \mathbf{r}\right) ;t\right\rangle $, where $\mathbf{r}$ denotes an
arbitrary field point within the system \cite{Origins}. These will be
referred to as the hydrodynamic fields since they are the ones that are
expected to obey the hydrodynamic equations under appropriate conditions.
The phase functions $m\left( \Gamma ,\mathbf{r}\right) ,e\left( \Gamma ,%
\mathbf{r}\right) ,$ and $\mathbf{g}\left( \Gamma ,\mathbf{r}\right) $ are
well known and their explicit forms will not be needed here. They will be
denoted collectively by $a_{\alpha }\left( \mathbf{r}\right) $%
\begin{equation}
a_{\alpha }(\mathbf{r},t)\leftrightarrow \left\{ m\left( \mathbf{r}\right)
,e(\mathbf{r}),\mathbf{g}(\mathbf{r})\right\} .  \label{4.1}
\end{equation}%
It follows from (\ref{3.15}) that they obey the microscopic balance
equations \cite{DBB07}%
\begin{equation}
\partial _{t}a_{\alpha }(\mathbf{r},t)=La_{\alpha }(\mathbf{r},t)=-\nabla
\cdot \mathbf{b}_{\alpha }(\mathbf{r},t)-\delta _{\alpha 2}w(\mathbf{r},t).
\label{4.2}
\end{equation}%
To obtain this result, it has been recognized that the quantity $La_{\alpha
}(\mathbf{r},t)$ can be written as the sum of a divergence $\nabla \cdot
\mathbf{b}_{\alpha }(\mathbf{r},t)$ plus a remainder $w(\mathbf{r},t)$ that
cannot be so represented. For a normal fluid $w(\mathbf{r},t)$ vanishes and (%
\ref{4.2}) become the local conservation laws for mass, energy, and
momentum. The $\mathbf{b}_{\alpha }(\mathbf{r},t)$ are the corresponding
fluxes. This clarifies why $a_{\alpha }(\mathbf{r},t)$ are selected for a
macroscopic description. Their time dependence is determined by the scale of
the spatial gradients of the fluxes, and averages of the latter become small
as the system approaches homogeneity. Consequently, on long time scales the $%
\left\langle a_{\alpha }\left( \mathbf{r}\right) ;t\right\rangle $ are the
only surviving dynamical variables, and it is under these conditions that
these fields obey hydrodynamic equations. The mass and momentum are
conserved for a granular fluid as well, but there is a loss of energy $w(%
\mathbf{r},t)$ due to the non-conservative forces. It is no longer obvious
that the energy is still one of the slow variables since its time scale is
coupled to $w(\mathbf{r},t)$ which does not become small for nearly
homogeneous states. Thus, an additional requirement for the existence of a
macroscopic description in terms of $\left\langle a_{\alpha }\left( \mathbf{r%
}\right) ;t\right\rangle $ is that the time scale of $\left\langle e\left(
\mathbf{r}\right) ;t\right\rangle /\left\langle w\left( \mathbf{r}\right)
;t\right\rangle $ must be larger than that for non-hydrodynamic properties.
This issue is discussed further below.

The macroscopic balance equations follow from the averages of (\ref{4.2})%
\begin{equation}
\partial _{t}\left\langle a_{\alpha }\left( \mathbf{r}\right)
;t\right\rangle +\nabla \cdot \left\langle \mathbf{b}_{\alpha }\left(
\mathbf{r}\right) ;t\right\rangle =-\delta _{\alpha 2}\left\langle w\left(
\mathbf{r}\right) ;t\right\rangle .  \label{4.3}
\end{equation}%
These equations are formally exact, but of little practical use as they do
not form a closed (self-deterministic) set of equations for $\left\langle
a_{\alpha }\left( \mathbf{r}\right) ;t\right\rangle $. Closure requires
expressing the average flux $\left\langle \mathbf{b}_{\alpha }\left( \mathbf{%
r}\right) ;t\right\rangle $ and energy loss $\left\langle w\left( \mathbf{r}%
\right) ;t\right\rangle $ as functionals of the fields $\left\langle
a_{\alpha }\left( \mathbf{r}\right) ;t\right\rangle .$ Such relationships
are called "constitutive equations". The combination of the exact balance
equations with some form of constitutive equations provides the most general
definition of hydrodynamics.

Construction of the constitutive equations is simplified by extracting the
effects of convection. The velocity $\mathbf{U}(\mathbf{r},t)$ of a cell at
point $\mathbf{r}$ is defined in terms of the average momentum
\begin{equation}
\left\langle \mathbf{g}\left( \mathbf{r}\right) ;t\right\rangle \equiv
\left\langle m\left( \mathbf{r}\right) ;t\right\rangle \mathbf{U}(\mathbf{r}%
,t).  \label{4.6}
\end{equation}%
The fluxes are functions of the positions and velocities $\mathbf{b}_{\alpha
}\left( \mathbf{r}\right) =\mathbf{b}_{\alpha }\left( \mathbf{r;}\{\mathbf{q}%
_{i},\mathbf{v}_{i}\}\right) =\mathbf{b}_{\alpha }\left( \mathbf{r;}\{%
\mathbf{q}_{i},\mathbf{V}_{i}+\mathbf{U(r)}\}\right) $, where the velocity
in the local rest frame has been introduced, $\mathbf{V}_{i}=\mathbf{v}_{i}-%
\mathbf{U}(\mathbf{r},t)$. Then defining the microscopic flux in the rest
frame by $\mathbf{b}_{\alpha }^{\prime }\left( \mathbf{r}\right) =\mathbf{b}%
_{\alpha }\left( \mathbf{r;}\{\mathbf{q}_{i},\mathbf{V}_{i}\}\right) $ it
follows that the average flux has the form \cite{McL}
\begin{equation}
\left\langle \mathbf{b}_{\alpha }\left( \mathbf{r}\right) ;t\right\rangle
=\left\langle \mathbf{b}_{\alpha }^{\prime }\left( \mathbf{r}\right)
;t\right\rangle +\mathbf{c}_{\alpha \eta }\mathbf{U}(\mathbf{r},t)\cdot
\left\langle \mathbf{b}_{\eta }^{\prime }\left( \mathbf{r}\right)
;t\right\rangle +\mathbf{U}(\mathbf{r},t)d_{\alpha }\left( \left\{
\left\langle a_{\nu }\left( \mathbf{r}\right) ;t\right\rangle \right\}
\right) .  \label{4.7}
\end{equation}%
The first term is the flux of mass, energy, and momentum in a fluid element
at rest, and represents the dissipative processes. The second and third
terms are proportional to the flow velocity $\mathbf{U}(\mathbf{r},t)$ and
are associated with convection. The coefficients of these terms are explicit
functions of the fields $\left\langle a_{\alpha }\left( \mathbf{r}\right)
;t\right\rangle $ (as is $\mathbf{U}(\mathbf{r},t)$\textbf{). } For a normal
fluid, neglect of the rest frame fluxes leads to the perfect fluid Euler
hydrodynamic equations. Hence, determination of the constitutive equations
is reduced to expressing the rest frame fluxes and energy loss as
functionals of the fields.

\section{"Normal" States and Hydrodynamics}

\label{sec5}

A hydrodynamic description is a closed set of equations for the hydrodynamic
fields, $\left\langle a_{\alpha }\left( \mathbf{r}\right) ;t\right\rangle .$
This follows from the exact macroscopic balance equations if the energy loss
and fluxes can be represented as functionals of these fields%
\begin{equation}
\left\langle w\left( \mathbf{r}\right) ;t\right\rangle \rightarrow \omega (%
\mathbf{r}\mid \left\langle a_{\alpha };t\right\rangle ),\hspace{0.3in}%
\left\langle \mathbf{b}_{\alpha }\left( \mathbf{r}\right) ;t\right\rangle
\rightarrow \mathbf{\beta }_{\alpha }(\mathbf{r}\mid \left\langle a_{\alpha
};t\right\rangle ).  \label{5.1}
\end{equation}%
The arrow is used to indicate that such a functional representation need not
be valid on all length and time scales, and any such restrictions constitute
the domain of validity for hydrodynamics. The notation here and below is
such that $f\left( \mathbf{r},t,\left\{ \left\langle a_{\alpha }\left(
\mathbf{r}\right) ;t\right\rangle \right\} \right) $ denotes a \emph{function%
} of $\mathbf{r},t$ and of the fields $\left\langle a_{\alpha }\left(
\mathbf{r}\right) ;t\right\rangle $ at the point $\mathbf{r}$, while $%
f\left( \mathbf{r},t\mid \left\langle a_{\alpha };t\right\rangle \right) $
denotes a function of $\mathbf{r},t$ and a \emph{functional} of the $%
\left\langle a_{\alpha };t\right\rangle $ at all space points. With such
constitutive relations the macroscopic balance equations (\ref{4.3}) become
hydrodynamic equations%
\begin{equation}
\partial _{t}\left\langle a_{\alpha }\left( \mathbf{r}\right)
;t\right\rangle +\nabla \cdot \mathbf{\beta }(\mathbf{r}\mid \left\langle
a_{\alpha };t\right\rangle )=-\delta _{\alpha 2}\omega (\mathbf{r}\mid
\left\langle a_{\alpha };t\right\rangle ).  \label{5.2}
\end{equation}

The average energy loss and fluxes are averages of specific functions of the
particle positions and velocities, and hence are linear functionals of the
solution to the Liouville equation. The existence of constitutive equations
is therefore related to a special property of the solution which will be
called "normal" (this terminology originates in a related context for
derivation of hydrodynamics from the Boltzmann kinetic equation \cite{McL}).
The class of \textquotedblright normal\textquotedblright\ distributions is
defined by the functional forms%
\begin{equation}
\rho _{n}\left( \Gamma ,t\right) =\rho _{n}\left( \{\mathbf{q}_{ij},\mathbf{v%
}_{i}\}\mid \left\langle a_{\alpha };t\right\rangle \right) .  \label{5.4}
\end{equation}%
All time dependence and all the breaking of translational invariance for
normal states occurs only through the hydrodynamic fields. A familiar
example of a normal distribution for real fluids is the \emph{local} Gibbs
distribution%
\begin{equation}
\rho _{e\ell }\left( \Gamma \mid \left\langle a_{\alpha };t\right\rangle
\right) =\exp \left\{ q-\int d\mathbf{r}y_{\alpha }\left( \mathbf{r}%
,t\right) a_{\alpha }\left( \mathbf{r}\right) \right\}  \label{5.4a}
\end{equation}%
Here $q$ is a normalization constant, and $y_{\alpha }\left( \mathbf{r}%
,t\right) $ are conjugate fields determined by the requirement that the
averages of $a_{\alpha }\left( \mathbf{r}\right) $ give the specified values
$\left\langle a_{\alpha }\left( \mathbf{r}\right) ;t\right\rangle $. In this
way $y_{\alpha }\left( \mathbf{r},t\right) $ are functionals of the
hydrodynamic fields and $\rho _{e\ell }\left( \Gamma \mid \left\langle
a_{\alpha };t\right\rangle \right) $ is normal. The importance of normal
solutions is that they yield directly the desired functionals of (\ref{5.1})
\begin{equation}
\omega (\mathbf{r}\mid \left\langle a_{\alpha };t\right\rangle )=\int
d\Gamma \rho _{n}\left( \Gamma \mid \left\langle a_{\alpha };t\right\rangle
\right) w(\mathbf{r})  \label{5.6}
\end{equation}%
\begin{equation}
\mathbf{\beta }_{\alpha }(\mathbf{r}\mid \left\langle a_{\alpha
};t\right\rangle )=\int d\Gamma \rho _{n}\left( \Gamma \mid \left\langle
a_{\alpha };t\right\rangle \right) \mathbf{b}_{\alpha }\left( \mathbf{r}%
\right)  \label{5.7}
\end{equation}

The normal state in (\ref{5.6}) and (\ref{5.7}) must be a solution to the
Liouville equation. In general, the time derivative in the Liouville
equation can be separated into that which occurs through $\left\langle
a_{\alpha };t\right\rangle $ plus the residual time dependence, generalizing
(\ref{3.17})%
\begin{equation}
\rho \left( \Gamma ,t\right) =\rho \left( \Gamma ,t\mid \left\langle
a_{\alpha };t\right\rangle \right) .  \label{5.8}
\end{equation}%
The Liouville equation then becomes%
\begin{equation}
\partial _{t}\rho \mid _{\left\langle a_{\alpha };t\right\rangle }-\int d%
\mathbf{r}\frac{\delta \rho }{\delta \left\langle a_{\alpha }\left( \mathbf{r%
}\right) ;t\right\rangle }\left\{ \nabla \cdot \left\langle \mathbf{b}%
_{\alpha }\left( \mathbf{r}\right) ;t\right\rangle +\delta _{\alpha
2}\left\langle w\left( \mathbf{r}\right) ;t\right\rangle \right\} +\overline{%
L}\rho =0.  \label{5.9}
\end{equation}%
A normal solution results when $\partial _{t}\rho _{n}\mid _{\left\langle
a_{\alpha };t\right\rangle }\rightarrow 0$. For specified fields, (\ref{5.9}%
) becomes an equation for the $\Gamma $ dependence of the normal phase space
density as a functional of the fields. This dependence then allows
determination of the normal forms in (\ref{5.6}) and (\ref{5.7}). Finally,
with the form of the hydrodynamic equation determined at that point, their
solution with suitable initial and boundary conditions provides the explicit
forms for the fields, and completes the normal solution. The existence and
determination of this solution is the central problem for establishing a
hydrodynamic description for both normal and granular fluids.

The concept of a normal solution and its use in the macroscopic balance
equations makes no special reference to whether the fluid is normal or
granular, and is not restricted to states near homogeneity. In this general
context, hydrodynamics is not a simple set of local partial differential
equations such as the familiar Navier-Stokes equations. The latter are a
special case of this more general idea, and their inadequacy for some
conditions should not be interpreted as the absence of a more complex
hydrodynamic description.

In closing this Section a qualitative explanation of why a normal solution
can be expected is given, by analogy with the similar expectation for atomic
fluids. For a wide class of initial states there is a first stage of rapid
velocity relaxation in each small region toward the universal homogeneous
state (HCS or Gibbs, respectively). However, the hydrodynamic parameters of
that universal state are specific to each region so it is only locally
homogenous, as in (\ref{5.4a}) for the atomic fluid. Subsequently, these
differences in the parameters of neighboring cells are decreased by the
fluxes of mass, energy, and momentum across their boundaries. It is this
second stage where a normal description in terms of the hydrodynamic fields
can be expected, indicating also that the space and time scales for a
hydrodynamic description should be large compared to those of the first
stage. This basic conceptual picture is essentially the same for both atomic
and granular fluids, and the rapid approach of the first stage is indeed
observed in molecular dynamics simulation studies of both equilibrium and
the HCS.

\section{Navier-Stokes Approximation}

\label{sec6}

Equation (\ref{5.9}) presents a formidable problem and further progress
requires specialization to specific cases of interest. Perhaps the simplest
of these are weakly inhomogeneous states. These are states for which all
spatial gradients of first order are small and all higher order derivatives
are negligible. Small gradients means that the relative change in the
hydrodynamic fields over the largest microscopic length scale $\ell _{0}$ is
small: $\ell _{0}\partial _{r}\ln \left\langle a_{\alpha };t\right\rangle
<<1.$ There are two characteristic length scales, the mean free path and the
grain diameter. For a dilute gas the mean free path is largest, while for a
dense fluid the grain size is largest. Under these conditions a solution to
the Liouville equation can be sought as an expansion to leading order in
these small gradients. This will be referred to as the Navier-Stokes
approximation.

According to the discussion at the end of the last section, a normal
solution is expected after the system has relaxed to its local HCS form,
denoted by $\rho _{0\ell }\left( \Gamma \mid \left\langle a_{\alpha
};t\right\rangle \right) $, representing the fluid as having each cell in
its own HCS. Define the deviations of the hydrodynamic fields from some
common reference value by%
\begin{equation}
\delta \left\langle a_{\alpha };t\right\rangle =\left\langle a_{\alpha
};t\right\rangle -a_{0\alpha },  \label{6.1}
\end{equation}%
where $a_{0\alpha }$ is the same for all cells. Then, the local HCS must
satisfy the conditions
\begin{equation}
\rho _{0\ell }\left( \Gamma \mid a_{0\alpha }+\delta \left\langle a_{\alpha
};t\right\rangle \right) \mid _{\delta \left\langle a_{\alpha
};t\right\rangle =0}=\rho _{0}\left( \Gamma ;a_{0\alpha }\right) ,
\label{6.3}
\end{equation}%
\begin{equation}
\frac{\partial \rho _{0}}{\partial a_{0\alpha }}=\int d\mathbf{r}\frac{%
\delta \rho _{0\ell }\left( \Gamma \mid a_{0\alpha }+\delta \left\langle
a_{\alpha };t\right\rangle \right) }{\delta \left\langle a_{\alpha }\left(
\mathbf{r}\right) ;t\right\rangle _{\alpha }}\mid _{\delta \left\langle
a_{\alpha };t\right\rangle =0},\cdot \cdot \cdot  \label{6.3a}
\end{equation}%
\ i.e., the local HCS and all of its functional derivatives must agree with
those of the HCS in the homogenous limit. Also, as a normal distribution its
time dependence is through the exact hydrodynamic fields for the fluid state
considered. This means the averages of the corresponding microscopic fields $%
a_{\alpha }\left( \mathbf{r}\right) $ for the local HCS and for the solution
to the Liouville equation must be the same
\begin{equation}
\int d\Gamma \left( \rho -\rho _{0\ell }\right) a_{\alpha }\left( \mathbf{r}%
\right) =0.  \label{6.2}
\end{equation}%
A more complete discussion of the construction of $\rho _{0\ell }$ from
knowledge of $\rho _{0}$ is given elsewhere \cite{DBB07}. For the purposes
here properties (\ref{6.3}), (\ref{6.3a}), and (\ref{6.2}) are sufficient.

The local HCS distribution, $\rho _{0\ell },$ is not a solution to the
Liouville equation except in limit that all hydrodynamic fields become the
same for each cell. Instead, it is a reference state approximating the
actual solution after its first stage of velocity relaxation. To construct a
solution $\rho $ define its deviation from $\rho _{0\ell }$ by
\begin{equation}
\rho \left( \Gamma ,t\mid \left\langle a_{\alpha };t\right\rangle \right)
=\rho _{0\ell }\left( \Gamma ,t\mid \left\langle a_{\alpha };t\right\rangle
\right) +\Delta \left( \Gamma ,t\mid \left\langle a_{\alpha };t\right\rangle
\right) .  \label{6.4}
\end{equation}%
The Liouville equation (\ref{5.9}) gives
\begin{eqnarray}
&&\partial _{t}\Delta -\int d\mathbf{r}^{\prime }\frac{\delta \Delta }{%
\delta \left\langle a_{\alpha }\left( \mathbf{r}^{\prime }\right)
;t\right\rangle }\left\{ \nabla \cdot \left\langle \mathbf{b}_{\alpha
}\left( \mathbf{r}\right) ;t\right\rangle +\delta _{\alpha 2}\left\langle
w\left( \mathbf{r}\right) ;t\right\rangle \right\} +\overline{L}\Delta
\notag \\
&=&\int d\mathbf{r}^{\prime }\frac{\delta \rho _{0\ell }}{\delta
\left\langle a_{\alpha }\left( \mathbf{r}^{\prime }\right) ;t\right\rangle }%
\left\{ \nabla \cdot \left\langle \mathbf{b}_{\alpha }\left( \mathbf{r}%
\right) ;t\right\rangle +\delta _{\alpha 2}\left\langle w\left( \mathbf{r}%
\right) ;t\right\rangle \right\} -\overline{L}\rho _{0\ell }.  \label{6.5}
\end{eqnarray}%
This equation is still exact, but if only small gradient states are
considered it simplifies by retaining terms only of first order in the
gradients. To be precise, the ultimate use of this solution is to calculate
local properties of the form%
\begin{equation}
A(\mathbf{r},t\mid \left\{ y_{\alpha }\left( t\right) \right\} )=\int
d\Gamma a(\Gamma ,\mathbf{r})\rho \left( \Gamma ,t\mid \left\langle
a_{\alpha }(\mathbf{r)};t\right\rangle \mathbf{+}\delta \left\langle
a_{\alpha };t\right\rangle \right) .  \label{6.6}
\end{equation}%
Therefore, in the following analysis the gradient expansions are referred to
the field point $\mathbf{r}$ of interest, $\left\langle a_{\alpha
};t\right\rangle =\left\langle a_{\alpha }(\mathbf{r)};t\right\rangle
\mathbf{+}\delta \left\langle a_{\alpha };t\right\rangle $, i.e. the common
reference values in (\ref{6.1}) are the exact field values at the chosen
point, $a_{0\alpha }=\left\langle a_{\alpha }(\mathbf{r)};t\right\rangle $.
The gradient expansion is carried out relative to these values. Of course
the results will be general and applicable to any choice for $\mathbf{r}$.

The details of the gradient expansion are given in the Appendix. The
solution to the Liouville equation to first order in the gradients is%
\begin{eqnarray}
\rho \left( \Gamma ,t\mid \left\langle a_{\alpha };t\right\rangle \right)
&=&\rho _{0}\left( \Gamma ,\left\langle a_{\alpha }(\mathbf{r)}%
;t\right\rangle \right) \mathbf{+}\left( 1-\mathcal{P}\right) \left( \mathbf{%
M}_{\beta }\left( \Gamma ,\left\langle a_{\alpha }(\mathbf{r)}%
;t\right\rangle \right) \right.  \notag \\
&&\left. +\int_{0}^{t}dt^{\prime }\left( e^{-\left( I\overline{\mathcal{L}}%
+K^{T}\right) t^{\prime }}\right) _{\beta \nu }\left( 1-\mathcal{P}\right)
\boldsymbol{\Upsilon }_{\nu }\left( \Gamma ,\left\langle a_{\alpha }(\mathbf{%
r)};t\right\rangle \right) \right) \cdot \nabla \left\langle a_{\beta
}\left( \mathbf{r}\right) ;t\right\rangle .  \notag \\
&&  \label{6.7}
\end{eqnarray}%
with the definitions%
\begin{equation}
\mathbf{M}_{\beta }=\int d\mathbf{r}^{\prime }\left( \frac{\delta \rho
_{0\ell }}{\delta \left\langle a_{\alpha }\left( \mathbf{r}^{\prime }\right)
;t\right\rangle }\right) _{\delta \left\langle a_{\alpha };t\right\rangle =0}%
\mathbf{r}^{\prime },  \label{6.8}
\end{equation}%
\begin{equation}
\boldsymbol{\Upsilon }_{\alpha }=-\left( I\overline{\mathcal{L}}%
+K^{T}\right) _{\alpha \beta }\mathbf{M}_{\beta }.  \label{6.9}
\end{equation}%
The generator for the dynamics $I\overline{\mathcal{L}}+K^{T}$ has a
contribution from $\overline{\mathcal{L}}$ which is the same as in (\ref%
{3.19}), with $\omega $ evaluated for the HCS as a function of the exact
hydrodynamic fields at the point $\mathbf{r}$ and time $t$
\begin{equation}
\overline{\mathcal{L}}=-\omega _{0}(\left\langle a_{\alpha }\left( \mathbf{r}%
\right) ;t\right\rangle )\partial _{\left\langle e\left( \mathbf{r}\right)
;t\right\rangle }+\overline{L}.  \label{6.10}
\end{equation}%
The second contribution to the generator of the dynamics is the transpose of
the matrix $K_{\alpha \beta }$%
\begin{equation}
K_{\alpha \beta }=\delta _{\alpha 2}\frac{\partial \omega _{0}(\left\langle
a_{\alpha }\left( \mathbf{r}\right) ;t\right\rangle )}{\partial \left\langle
a_{\beta }\left( \mathbf{r}\right) ;t\right\rangle }.  \label{6.11}
\end{equation}%
Finally, $\mathcal{P}$ is a projection operator%
\begin{equation}
\mathcal{P}X=\Psi _{\beta }\int d\Gamma A_{\beta }X,\hspace{0.3in}A_{\beta
}=V^{-1}\int d\mathbf{r}a_{\alpha }\left( \mathbf{r}\right) ,\hspace{0.3in}%
\Psi _{\beta }\equiv \frac{\partial \rho _{0}}{\partial \left\langle
a_{\beta }\left( \mathbf{r}\right) ;t\right\rangle }  \label{6.12}
\end{equation}%
The phase functions $A_{\beta }$ and $\Psi _{\beta }$ form a biorthogonal
set in the sense%
\begin{equation}
\int d\Gamma A_{\alpha }\Psi _{\beta }=\delta _{\alpha \beta }.  \label{6.13}
\end{equation}%
The $A_{\alpha }$ are the usual global invariants of the Liouville operator $%
\overline{L}$ for a normal fluid; it is shown in the Appendix that the $\Psi
_{\beta }$ are the invariants of the new generator for dynamics in a
granular fluid
\begin{equation}
\left( I\overline{\mathcal{L}}_{T}+K^{T}\right) _{\nu \beta }\Psi _{\beta
}=0.  \label{6.14}
\end{equation}

Equation (\ref{6.7}) is not quite the normal solution desired. All terms
depend on time through\ $\left\langle a_{\beta }\left( \mathbf{r}\right)
;t\right\rangle $ as required, except for the last term which has an
additional explicit time dependence through the upper limit of the time
integral. This time dependence becomes negligible if the integrand is
effectively non-zero after some short time scale $\tau $. Then for $t>>\tau $
the time integral becomes independent of $t$ and can be taken formally to
infinity. Thus, a normal solution is attained for this time scale%
\begin{eqnarray}
\rho _{n}\left( \Gamma ,\left\langle a_{\alpha }\left( \mathbf{r}\right)
;t\right\rangle \right) &=&\rho _{0}\left( \Gamma ,\left\langle a_{\alpha }(%
\mathbf{r)};t\right\rangle \right) \mathbf{+}\left( 1-\mathcal{P}\right)
\left( \mathbf{M}_{\beta }\left( \Gamma ,\left\langle a_{\alpha }(\mathbf{r)}%
;t\right\rangle \right) \right.  \notag \\
&&\left. +\lim_{t_{0}\rightarrow \infty }\int_{0}^{t_{0}}dt^{\prime }\left(
e^{-\left( I\overline{\mathcal{L}}+K^{T}\right) t^{\prime }}\right) _{\beta
\nu }\left( 1-\mathcal{P}\right) \boldsymbol{\Upsilon }_{\nu }\left( \Gamma
,\left\langle a_{\alpha }(\mathbf{r)};t\right\rangle \right) \right) \cdot
\nabla \left\langle a_{\beta }\left( \mathbf{r}\right) ;t\right\rangle .
\notag \\
&&  \label{6.15}
\end{eqnarray}%
It is expected that the integrand should have this property of a short time
scale since the domain of operation for the generator of time dependence is
functions with translational invariance (as a consequence of the gradient
expansion). Hence there are no explicit slow hydrodynamic modes of finite
wavelength. Also, there is no contribution from the homogeneous
hydrodynamics (that for the invariants) due to the orthogonal projection $%
\left( 1-\mathcal{P}\right) $. The appearance of this projection is an
essential self-consistency of the analysis, and occurs as well for normal
fluids. The expression (\ref{6.15}) is only formal and the actual limit
should be taken in the weak sense only after (\ref{6.7}) has been used to
define average properties. A technical complication is the occurrence of
periodic time dependence, the Poincare recurrence time. This can be removed
by considering the thermodynamic limit of $V\rightarrow \infty ,N\rightarrow
\infty $ at constant $N/V$. Therefore, averages using the normal solution to
the Liouville equation are understood as having the thermodynamic limit
followed by the long time limit at constant $\left\langle a_{\alpha }(%
\mathbf{r)};t\right\rangle $.

An alternative equivalent form results from performing the integral in (\ref%
{6.15}) using the explicit form (\ref{6.9}) and the property (\ref{a.21}) of
the Appendix
\begin{eqnarray}
\rho _{n}\left( \Gamma ,\left\langle a_{\alpha }\left( \mathbf{r}\right)
;t\right\rangle \right) &=&\rho _{0}\left( \Gamma ,\left\langle a_{\alpha }(%
\mathbf{r)};t\right\rangle \right)  \notag \\
&&\mathbf{+}\lim_{t_{0}\rightarrow \infty }\left( 1-\mathcal{P}\right)
\left( e^{-\left( I\overline{\mathcal{L}}+K^{T}\right) t_{0}}\right) _{\beta
\nu }\mathbf{M}_{\nu }\left( \Gamma ,\left\langle a_{\alpha }(\mathbf{r)}%
;t\right\rangle \right) \cdot \nabla \left\langle a_{\beta }\left( \mathbf{r}%
\right) ;t\right\rangle .  \notag \\
&&  \label{6.16}
\end{eqnarray}%
The decay time for the integrand of (\ref{6.15}) now becomes the time after
which (\ref{6.16}) reaches its normal form.

\subsection{Constitutive equations}

The exact macroscopic balance equations are given by (\ref{5.2}), and the
necessary constitutive equations are given by (\ref{5.6}) and (\ref{5.7}) as
averages over the normal solution. These can be made more explicit now using
the small gradient result (\ref{6.15}). Since the latter is a local function
of the fields, the constitutive equations also will be local. Furthermore,
since all components of the gradients in (\ref{6.16}) depend on the common
value $\left\langle \mathbf{g}\left( \mathbf{r}\right) ;t\right\rangle $,
this can be eliminated through a Galilean transformation so that all
properties refer to a fluid element at rest. Of course, the gradients of $%
\left\langle \mathbf{g}\left( \mathbf{r}\right) ;t\right\rangle $ in that
fluid element are nonzero.

Consider first the energy loss function $\omega $
\begin{equation}
\omega (\left\langle a_{\alpha }\left( \mathbf{r}\right) ;t\right\rangle
)=\int d\Gamma \rho _{n}\left( \Gamma ,\left\langle a_{\alpha }\left(
\mathbf{r}\right) ;t\right\rangle \right) w(\mathbf{r})=\int d\Gamma \rho
_{n}\left( \Gamma ,\left\langle a_{\alpha }\left( \mathbf{r}\right)
;t\right\rangle \right) \overline{w}.  \label{6.17}
\end{equation}%
The coefficients of the gradient in the normal solution have translational
invariance and the average is independent of $\mathbf{r}$, except through
its parameterization by $\left\langle a_{\alpha }\left( \mathbf{r}\right)
;t\right\rangle $. The second equality takes this into account by replacing $%
w(\mathbf{r})$ by its average $\overline{w}$
\begin{equation}
\overline{w}=V^{-1}\int d\mathbf{r}w(\mathbf{r}).  \label{6.18}
\end{equation}%
Since $\omega (\left\langle a_{\alpha }\left( \mathbf{r}\right)
;t\right\rangle )$ is a scalar, fluid symmetry restricts the contributions
to first order in the gradients to
\begin{equation}
\omega (\left\langle a_{\alpha }\left( \mathbf{r}\right) ;t\right\rangle
)=\omega _{0}(\left\langle a_{\alpha }\left( \mathbf{r}\right)
;t\right\rangle )+\omega _{1}(\left\langle a_{\alpha }\left( \mathbf{r}%
\right) ;t\right\rangle )\nabla \cdot \mathbf{U}\left( \mathbf{r},t\right) .
\label{6.19}
\end{equation}%
Here, the flow velocity $\mathbf{U}\left( \mathbf{r},t\right) $ of (\ref{4.6}%
) has been used in place of the momentum density. The first term is the
contribution from the HCS distribution%
\begin{equation}
\omega _{0}(\left\langle a_{\alpha }\left( \mathbf{r}\right) ;t\right\rangle
)=\int d\Gamma \rho _{0}\left( \Gamma ,\left\langle a_{\alpha }\left(
\mathbf{r}\right) ;t\right\rangle \right) \overline{w}.  \label{6.20}
\end{equation}%
The coefficient of $\nabla \cdot \mathbf{U}\left( \mathbf{r},t\right) $ is
\begin{equation}
\omega _{1}(\left\langle a_{\alpha }\left( \mathbf{r}\right) ;t\right\rangle
)=\lim_{t_{0}\rightarrow \infty }C_{\omega }(t_{0},\left\langle a_{\alpha
}\left( \mathbf{r}\right) ;t\right\rangle )=C_{\omega }(0,\left\langle
a_{\alpha }\left( \mathbf{r}\right) ;t\right\rangle )+\lim_{t_{0}\rightarrow
\infty }\int_{0}^{t_{0}}\partial _{t^{\prime }}C_{\omega }(t^{\prime
},\left\langle a_{\alpha }\left( \mathbf{r}\right) ;t\right\rangle
)dt^{\prime }  \label{6.21}
\end{equation}%
with the correlation function defined by%
\begin{equation}
C_{\omega }(t)=\int d\Gamma \overline{w}\left( 1-\mathcal{P}\right) e^{-%
\overline{\mathcal{L}}t}M_{U},  \label{6.22}
\end{equation}%
\begin{equation}
M_{U}\equiv \frac{1}{3}\int d\mathbf{r}^{\prime }\mathbf{r}^{\prime }\cdot
\left( \frac{\delta \rho _{0\ell }}{\delta \mathbf{U}\left( \mathbf{r}%
^{\prime },t\right) }\right) _{\delta \left\langle a_{\alpha
};t\right\rangle =0}.  \label{6.23}
\end{equation}

The coefficient $\omega _{0}$ defines an "equation of state" for the
granular hydrodynamics, and gives the first non-trivial result of this
analysis. It is similar to the pressure (given below) and is an inherent
property of the local state of each cell, independent of the gradients
between cells. In contrast, $\omega _{1}$ is a true transport coefficient
characterizing communication between cells. The first equality of (\ref{6.21}%
) provides the Helfand form for this coefficient, while the second equality
gives the equivalent Green-Kubo form. Each has its practical utility,
depending on the method used for its approximate evaluation. Both forms have
proven useful for normal fluids, and further discussion is provided below.
Both $\omega _{0}$ and the transport coefficient $\omega _{1}$ vanish for
normal fluids since they characterize collisional energy loss.

The fluxes $\mathbf{\beta }_{\alpha }$ of (\ref{5.7}) can be determined in a
similar way. \ As indicated in (\ref{4.7}), only the rest frame flux $%
\mathbf{\beta }_{\alpha }^{\prime }$ is required. Furthermore, since all
components of the gradients in (\ref{6.16}) depend on the common value $%
\left\langle \mathbf{g}\left( \mathbf{r}\right) ;t\right\rangle $, this can
be eliminated through a Galilean transformation so that all properties refer
to a fluid element at rest. Of course, the gradients of $\left\langle
\mathbf{g}\left( \mathbf{r}\right) ;t\right\rangle $ in that fluid element
are nonzero. The component $\mathbf{\beta }_{1}^{\prime }$ is the rest frame
mass flux which is expected to vanish in order to give the continuity
equation. This follows from the fact that $\mathbf{b}_{1}^{\prime }\left(
\mathbf{r}\right) $\textbf{\ }is the microscopic momentum density%
\begin{equation}
\mathbf{\beta }_{1}^{\prime }=\int d\Gamma \mathbf{g}\left( \mathbf{r}%
\right) \rho _{0}\left( \Gamma ,\left\langle a_{\alpha }(\mathbf{r)}%
;t\right\rangle \right) \mathbf{+}\lim_{t_{0}\rightarrow \infty }\int
d\Gamma \mathbf{g}\left( \mathbf{r}\right) \left( 1-\mathcal{P}\right)
\left( \cdot \cdot \cdot \right) =0  \label{6.24}
\end{equation}%
The first term vanishes since $\left\langle \mathbf{g}\left( \mathbf{r}%
\right) ;t\right\rangle =0$ in the rest frame, and the second term vanishes
since $\left( 1-\mathcal{P}\right) $ projects orthogonal to the mass,
energy, and momentum. Thus, the expected continuity equation is verified.

The fluxes $\mathbf{\beta }_{2}^{\prime }$ and $\mathbf{\beta }_{\alpha
}^{\prime }$ for $\alpha =3,4,5$ are the rest frame energy and momentum
fluxes. The energy flux transforms like a vector and therefore fluid
symmetry (translational and rotational invariance) requires that it can
depend only on gradients of scalars%
\begin{equation}
\mathbf{\beta }_{2}^{\prime }=-\lambda \left( \left\langle a_{\alpha }(%
\mathbf{r)};t\right\rangle \right) \nabla T(\mathbf{r},t\mathbf{)}-\mu
\left( \left\langle a_{\alpha }(\mathbf{r)};t\right\rangle \right) \nabla
\left\langle m\left( \mathbf{r}\right) ;t\right\rangle .  \label{6.25}
\end{equation}%
To make the connection with Fourier's law for an atomic fluid, a temperature
$T(\mathbf{r},t\mathbf{)}$ has been introduced through the definition%
\begin{equation}
\left\langle e\left( \mathbf{r}\right) ;t\right\rangle \equiv
e_{0}(\left\langle m\left( \mathbf{r}\right) ;t\right\rangle \mathbf{,}T(%
\mathbf{r},t\mathbf{)).}  \label{6.26}
\end{equation}%
For an atomic fluid the function $e_{0}(\left\langle m\left( \mathbf{r}%
\right) ;t\right\rangle \mathbf{,}T(\mathbf{r},t\mathbf{))}$ is chosen to be
the thermodynamic internal energy density. As there is no thermodynamics for
a granular fluid this function is arbitrary and simply constitutes a change
of variables from $\left\langle e\left( \mathbf{r}\right) ;t\right\rangle
,\left\langle m\left( \mathbf{r}\right) ;t\right\rangle $ to $T(\mathbf{r},t%
\mathbf{))},\left\langle m\left( \mathbf{r}\right) ;t\right\rangle \mathbf{.}
$ In this form (\ref{6.25}) is a generalization of Fourier's law where $%
\lambda $ is the thermal conductivity \cite{DuftyGubbins}. The contribution
from the gradient of the mass density is new to granular fluids ($\mu =0$
for atomic fluids). These coefficients are given by%
\begin{equation}
\lambda \left( \left\langle a_{\alpha }(\mathbf{r)};t\right\rangle \right)
=\lim_{t_{0}\rightarrow \infty }C_{\lambda }(t_{0},\left\langle a_{\alpha
}\left( \mathbf{r}\right) ;t\right\rangle )=C_{\lambda }(0,\left\langle
a_{\alpha }\left( \mathbf{r}\right) ;t\right\rangle )+\lim_{t_{0}\rightarrow
\infty }\int_{0}^{t_{0}}\partial _{t^{\prime }}C_{\lambda }(t^{\prime
},\left\langle a_{\alpha }\left( \mathbf{r}\right) ;t\right\rangle
)dt^{\prime }  \label{6.27}
\end{equation}%
\begin{equation}
\mu \left( \left\langle a_{\alpha }(\mathbf{r)};t\right\rangle \right)
=\lim_{t_{0}\rightarrow \infty }C_{\mu }(t_{0},\left\langle a_{\alpha
}\left( \mathbf{r}\right) ;t\right\rangle )=C_{\mu }(0,\left\langle
a_{\alpha }\left( \mathbf{r}\right) ;t\right\rangle )+\lim_{t_{0}\rightarrow
\infty }\int_{0}^{t_{0}}\partial _{t^{\prime }}C_{\mu }(t^{\prime
},\left\langle a_{\alpha }\left( \mathbf{r}\right) ;t\right\rangle
)dt^{\prime }  \label{6.28}
\end{equation}%
with the correlation functions%
\begin{equation}
C_{\lambda }(t)=\frac{1}{3}\int d\Gamma \mathbf{\beta }_{2}^{\prime }\cdot
\left( 1-\mathcal{P}\right) e^{-\left( \overline{\mathcal{L}}+K_{22}\right)
t}\mathbf{M}_{T},  \label{6.29}
\end{equation}%
\begin{eqnarray}
C_{\mu }(t) &=&\frac{1}{3}\int d\Gamma \mathbf{\beta }_{2}^{\prime }\cdot
\left( 1-\mathcal{P}\right) \left( e^{-\overline{\mathcal{L}}t}\left(
\mathbf{M}_{m}+\frac{K_{21}}{K_{22}}\mathbf{M}_{T}\right) \right.  \notag \\
&&+\left. e^{-\left( \overline{\mathcal{L}}+K_{22}\right) t}\left( \frac{%
\partial e_{0}}{\partial \left\langle m\left( \mathbf{r}\right)
;t\right\rangle }-\frac{K_{21}}{K_{22}}\right) \frac{\partial T}{\partial
e_{0}}\mid _{\left\langle m\left( \mathbf{r}\right) ;t\right\rangle }\mathbf{%
M}_{T}\right) ,  \label{6.30}
\end{eqnarray}%
\begin{equation}
\mathbf{M}_{T}\equiv \int d\mathbf{r}^{\prime }\mathbf{r}^{\prime }\left(
\frac{\delta \rho _{0\ell }}{\delta T\left( \mathbf{r}^{\prime },t\right) }%
\right) _{\delta \left\langle a_{\alpha };t\right\rangle =0}.  \label{6.32}
\end{equation}%
\begin{equation}
\mathbf{M}_{m}\equiv \int d\mathbf{r}^{\prime }\mathbf{r}^{\prime }\left(
\frac{\delta \rho _{0\ell }}{\delta \left\langle m\left( \mathbf{r}^{\prime
}\right) ;t\right\rangle }\right) _{\delta \left\langle a_{\alpha
};t\right\rangle =0}.  \label{6.31}
\end{equation}

Finally, the set of vectors $\mathbf{\beta }_{\alpha }^{\prime }$ for $%
\alpha =3,4,5$ define the pressure tensor $\mathbf{\beta }_{\alpha }^{\prime
}\Leftrightarrow P_{ij}.$ Fluid symmetry then determines that it can couple
only to the momentum gradients, or equivalently the flow velocity gradients,
in the form
\begin{eqnarray}
P_{ij} &=&p\left( \left\langle a_{\alpha }(\mathbf{r)};t\right\rangle
\right) \delta _{ij}-\eta \left( \left\langle a_{\alpha }(\mathbf{r)}%
;t\right\rangle \right) \left( \partial _{i}U_{j}\left( \mathbf{r},t\right)
+\partial _{j}U_{i}\left( \mathbf{r},t\right) -\frac{2}{3}\delta _{ij}%
\mathbf{\nabla \cdot U}\left( \mathbf{r},t\right) \right)  \notag \\
&&-\kappa \left( \left\langle a_{\alpha }(\mathbf{r)};t\right\rangle \right)
\delta _{ij}\mathbf{\nabla \cdot U}\left( \mathbf{r},t\right) .  \label{6.33}
\end{eqnarray}%
The scalar function $p\left( \left\langle a_{\alpha }(\mathbf{r)}%
;t\right\rangle \right) $ is the pressure, now identified as%
\begin{equation}
p\left( \left\langle a_{\alpha }(\mathbf{r)};t\right\rangle \right) =\int
d\Gamma \rho _{0}\left( \Gamma ,\left\langle a_{\alpha }\left( \mathbf{r}%
\right) ;t\right\rangle \right) \beta _{3x}^{\prime }.  \label{6.34}
\end{equation}%
The transport coefficients in (\ref{6.33}) are the shear viscosity $\eta
\left( \left\langle a_{\alpha }(\mathbf{r)};t\right\rangle \right) $ and the
bulk viscosity $\kappa \left( \left\langle a_{\alpha }(\mathbf{r)}%
;t\right\rangle \right) $ given by%
\begin{equation}
\eta \left( \left\langle a_{\alpha }(\mathbf{r)};t\right\rangle \right)
=\lim_{t_{0}\rightarrow \infty }C_{\eta }(t_{0},\left\langle a_{\alpha
}\left( \mathbf{r}\right) ;t\right\rangle )=C_{\eta }(0,\left\langle
a_{\alpha }\left( \mathbf{r}\right) ;t\right\rangle )+\lim_{t_{0}\rightarrow
\infty }\int_{0}^{t_{0}}\partial _{t^{\prime }}C_{\eta }(t^{\prime
},\left\langle a_{\alpha }\left( \mathbf{r}\right) ;t\right\rangle
)dt^{\prime }  \label{6.35}
\end{equation}%
\begin{equation}
\kappa \left( \left\langle a_{\alpha }(\mathbf{r)};t\right\rangle \right)
=\lim_{t_{0}\rightarrow \infty }C_{\kappa }(t_{0},\left\langle a_{\alpha
}\left( \mathbf{r}\right) ;t\right\rangle )=C_{\kappa }(0,\left\langle
a_{\alpha }\left( \mathbf{r}\right) ;t\right\rangle )+\lim_{t_{0}\rightarrow
\infty }\int_{0}^{t_{0}}\partial _{t^{\prime }}C_{\kappa }(t^{\prime
},\left\langle a_{\alpha }\left( \mathbf{r}\right) ;t\right\rangle
)dt^{\prime }  \label{6.36}
\end{equation}%
with the correlation functions%
\begin{equation}
C_{\eta }(t)=\int d\Gamma \beta _{3y}^{\prime }\cdot \left( 1-\mathcal{P}%
\right) e^{-\overline{\mathcal{L}}t}M_{\eta },  \label{6.37}
\end{equation}%
\begin{equation}
C_{\kappa }(t)=\int d\Gamma \beta _{3x}^{\prime }\cdot \left( 1-\mathcal{P}%
\right) e^{-\overline{\mathcal{L}}t}M_{\kappa },  \label{6.38}
\end{equation}%
\begin{equation}
M_{\eta }\equiv \int d\mathbf{r}^{\prime }x^{\prime }\left( \frac{\delta
\rho _{0\ell }}{\delta U_{y}\left( \mathbf{r}^{\prime },t\right) }\right)
_{\delta \left\langle a_{\alpha };t\right\rangle =0}.  \label{6.39}
\end{equation}%
\begin{equation}
M_{\kappa }\equiv \int d\mathbf{r}^{\prime }y^{\prime }\left( \frac{\delta
\rho _{0\ell }}{\delta U_{y}\left( \mathbf{r}^{\prime },t\right) }\right)
_{\delta \left\langle a_{\alpha };t\right\rangle =0}.  \label{6.40}
\end{equation}

This completes the formal derivation of the constitutive equations leading
to the nonlinear Navier-Stokes equations, including expressions for the
cooling rate, energy flux, and pressure tensor including contributions up
through first order in the gradients of the hydrodynamic fields. These
expressions are functions of the hydrodynamic fields to be determined by
their detailed many-body analysis of the correlation functions.

\subsection{Green-Kubo expressions}

To contrast the results here with those for an atomic fluid, it is
instructive to focus on the Green-Kubo forms for the transport coefficients
\cite{McL}. These are given by the second equalities of (\ref{6.21}), (\ref%
{6.27}), (\ref{6.28}), (\ref{6.35}), and (\ref{6.36}); the first equalities
are the corresponding Helfand forms \cite{Helfand}. For an atomic fluids
there is no counter part to $\omega _{1}$ and $\mu $. However, there are
Green-Kubo expressions for the thermal conductivity and the two viscosities.
For the discussion here only the thermal conductivity is considered, whose
Green-Kubo expression is%
\begin{equation}
\lambda \left( \left\langle a_{\alpha }(\mathbf{r)};t\right\rangle \right)
=C_{\lambda }(0,\left\langle a_{\alpha }\left( \mathbf{r}\right)
;t\right\rangle )+\lim_{t_{0}\rightarrow \infty }\int_{0}^{t_{0}}\partial
_{t^{\prime }}C_{\lambda }(t^{\prime },\left\langle a_{\alpha }\left(
\mathbf{r}\right) ;t\right\rangle )dt^{\prime }  \label{6.41}
\end{equation}%
\begin{equation}
\partial _{t}C_{\lambda }(t)=\frac{1}{3}\int d\Gamma \mathbf{\beta }%
_{2}^{\prime }\cdot \left( 1-\mathcal{P}\right) e^{-\left( \overline{%
\mathcal{L}}+K_{22}\right) t}\boldsymbol{\Upsilon }_{\lambda },  \label{6.42}
\end{equation}%
\begin{equation}
\boldsymbol{\Upsilon }_{e}=-\left( \overline{\mathcal{L}}+K_{22}\right)
\mathbf{M}_{e}  \label{6.43}
\end{equation}%
In contrast the thermal conductivity for an atomic fluid is%
\begin{equation}
\lambda \left( \left\langle a_{\alpha }(\mathbf{r)};t\right\rangle \right)
\rightarrow \lim_{t_{0}\rightarrow \infty }\int_{0}^{t_{0}}\partial
_{t^{\prime }}C_{\lambda }(t^{\prime },\left\langle a_{\alpha }\left(
\mathbf{r}\right) ;t\right\rangle )dt^{\prime }  \label{6.44}
\end{equation}%
\begin{equation}
\partial _{t}C_{\lambda }(t)\rightarrow \frac{1}{3T^{2}}\int d\Gamma \mathbf{%
\beta }_{2}^{\prime }\cdot \left( 1-\mathcal{P}\right) e^{-Lt}\mathbf{\beta }%
_{2}^{\prime }\rho _{e},  \label{6.45}
\end{equation}%
In this last expression $\rho _{e}$ is the equilibrium Gibbs ensemble, and
it is understood that $\mathbf{\beta }_{2}^{\prime }$ is the microscopic
expression for the energy flux for a dynamics with conservative forces, and $%
L$ is the Liouville generator for the corresponding dynamics.

There are several similarities and differences between the granular and
atomic fluid expressions \cite{DBB07}. The latter is the time integral of a
energy flux - energy flux equilibrium time correlation function. The
granular fluid is similar, with one of the fluxes the same but the other
flux is generated from the local HCS state. Also, the generator for the
dynamics in the granular case has two additional effects, $L$ replaced by $%
\overline{\mathcal{L}}+K_{22}$, to represents homogeneous cooling of the
reference state and its homogeneous response to perturbations. The
projection orthogonal to the invariants of each dynamics $\left( 1-\mathcal{P%
}\right) $ occurs in both cases as a necessary condition for the long time
limit of the time integral, and the corresponding existence of the normal
state. Finally, the contribution from $C_{\lambda }(0,\left\langle a_{\alpha
}\left( \mathbf{r}\right) ;t\right\rangle )$ vanishes for a normal fluid but
is non-zero for the granular fluid due to the non-conservative forces.

\subsection{Navier-Stokes hydrodynamic equations}

In closing this section it is appropriate to record the results of
substituting the Navier-Stokes constitutive equations, valid to first order
in the gradients, into the exact macroscopic balance equations. This defines
the Navier-Stokes hydrodynamic equations for a granular fluid%
\begin{equation}
D_{t}m+m\nabla _{\mathbf{r}}\cdot \mathbf{U}=0  \label{6.46}
\end{equation}%
\begin{equation*}
D_{t}e_{0}+\omega _{0}+\left( p+\omega _{1}+\left( \frac{2}{3}\eta -\kappa
\right) \nabla \cdot \mathbf{U}\right) \nabla \cdot \mathbf{U}
\end{equation*}%
\begin{equation}
-\eta \left( \partial _{\alpha }U_{\beta }+\partial _{\beta }U_{\alpha
}\right) \partial _{\alpha }U_{\beta }-\boldsymbol{\nabla }\cdot \left(
\lambda \mathbf{\nabla }T+\mu \boldsymbol{\nabla }m\right) =0,  \label{6.47}
\end{equation}%
\begin{equation}
D_{t}U_{\alpha }+m^{-1}\partial _{\alpha }\left( p-\left( \frac{2}{3}\eta
+\kappa \right) \nabla \cdot \mathbf{U}\right) -m^{-1}\partial _{\beta }\eta
\left( \partial _{\alpha }U_{\beta }+\partial _{\beta }U_{\alpha }\right) =0.
\label{6.48}
\end{equation}%
For simplicity of notation, $m\equiv \left\langle m\left( \mathbf{r}\right)
;t\right\rangle $ in these equations. They are a set of five nonlinear
partial differential equations for the variables $m,e_{0},$ and $\mathbf{U.}$
They are a closed set of equations since $\omega _{0}$, $p$, and the
transport coefficients $\omega _{1}$, $\lambda $, $\mu $, $\eta $, and $%
\kappa $ a defined as functions of these variables. These definitions for
the constitutive equations are the primary accomplishment of the statistical
mechanical basis for the hydrodynamic equations. The form of (\ref{6.35})-(%
\ref{6.36}) could have been guessed from the outset based on the macroscopic
balance equations and fluid symmetry. The underlying basis in the
microdynamics of the particles provides the necessary details for how the
parameters of these equations must depend on the fields. Here only the
formal definitions have been identified. It is only the first half of the
problem of completing these equations, as the evaluation of these
definitions poses a difficult many body problem. Still, without this first
half, the starting point for that detailed analysis would not be possible.
This the case for atomic fluids as well.

\section{Future directions}

\label{sec9}

The objective here has been to formulate the basis for a macroscopic
description of granular fluids using the fundamental principles of
nonequilibrium statistical mechanics. The analysis presented follows that
for an atomic fluid. First, the exact macroscopic balance equations are
identified. Next, their closure is linked to the concept of a normal state
and corresponding normal solution to the Liouville equation. This defines
the domain of hydrodynamics in its most general sense, both for atomic and
granular fluids. The construction of a normal solution is quite difficult in
general, but can be accomplished for states with small gradients relative to
locally homogeneous conditions. This gives the Navier-Stokes approximation
described here.

Navier-Stokes hydrodynamics is applicable for most common states of atomic
fluids, while deviations occur primarily for more complex polymeric
molecular fluids. The latter have rheological properties corresponding to
larger gradients relative to additional microscopic length and time scales.
The construction of normal states in these cases is more difficult and is
still at the semi-phenomenological stage \cite{Bird}. Granular fluids
provide a new motivation for renewed efforts to describe these more complex
normal states. The reason is that even structurally simple granular fluids
composed of spherically symmetric particles can exhibit rheology and other
phenomena beyond the Navier-Stokes domain of validity \cite{Inherent,Hrenya}%
. This is due to the cooling rate in the energy balance equation which
provides a new internal time scale, that can set the size of hydrodynamic
gradients beyond any control through boundary conditions. For example, new
steady states are possible for granular fluids due to the balance of this
internal cooling with external forcing. In many cases this implies that the
hydrodynamic description required is beyond the Navier-Stokes domain. The
understanding of constitutive equations in these cases is poor at this
point. It is hoped that the formal structure described here will provide the
appropriate basis for studies of these problems.

The context of hydrodynamics depends on the formation of a normal state from
more complex conditions. Above this has been described qualitatively as a
two stage process of rapid velocity relaxation in each cell to a state near
the local HCS, followed by hydrodynamic relaxation through exchange of mass,
energy, and momentum between the cells on a longer time scale. This
separation of microscopic and hydrodynamic time scales is essential to the
dominance of the hydrodynamic excitations over all others at large space and
time scales. It is justified for atomic fluids since the hydrodynamic times
are determined by the wavelength of the phenomena studied. As the system
approaches homogeneity, these time scales become much larger than the
microscopic excitations and hydrodynamics prevails at large times. However,
there is an additional hydrodynamic time scale for granular fluids, the
cooling rate, which is not set by the wavelength alone. It would seem that
this additional time scale must be large as well, implying a weak cooling
rate. This condition is too strong. What matters is the rate of the approach
to the homogeneous state, not any dynamics of that final state. In the above
derivation of hydrodynamics the final form for the solution to the Liouville
equation, eq. (\ref{6.15}) or (\ref{6.16}), has a dynamics generated by $I%
\overline{\mathcal{L}}+K^{T}$ rather than simply that for the trajectories $%
\overline{L}$. This is significant since the former has the additional
compensation for the cooling and for the homogeneous perturbations of that
cooling. Hence the approach to the time dependent normal state is determined
only by the remaining non-hydrodynamic relaxation. The time scale for
relaxation to the normal state is independent of the hydrodynamic time
scales of that normal state. Quantitative verification of these concepts is
another important future direction for research on a hydrodynamic
description for granular fluids.

\section{Acknowledgments}

The author is indebted to Professor J. Javier Brey and Dr. Aparna Baskaran
of Syracuse University for their collaboration on closely related linear
response methods for granular fluids. This manuscript has been prepared for
publication in the Encyclopedia of Complexity and Systems Science \cite%
{Encyclopedia}.

\section{Bibliography}

\noindent\textbf{Primary Literature}

\vspace{0.3in}

\noindent \textbf{Books and Reviews}

\noindent Campbell CS (1990) Rapid granular flows. Ann. Rev. Fluid Mech.%
\textit{,} Vol 22, pp 57-92

\noindent(1993) \emph{Granular Matter, An Interdisciplinary Approach}, Mehta
A, editor (Springer, NY)

\noindent Jaeger HM, Nagel SR, Behringer RP (1996) Granular solids, liquids,
and gases. Rev. Mod. Phys., Vol 68, pp 1259-1273.

\noindent Duran J (2000) Sands, powders, and grains: an introduction to the
physics of granular materials (Springer, NY).

\noindent (2001) \emph{Granular Gases}, P\"{o}schel T, Luding S, eds.
(Springer, NY).

\noindent (2002) \emph{Challenges in Granular Physics}, Halsey T, Metha A,
eds., (World Scientific, Singapore).

\noindent Campbell CS (2002) Granular shear flows in the elastic limit. J.
Fluid Mech., Vol 465, pp 261-291.

\noindent (2003) \emph{Granular Gases Dynamics}, P\"{o}schel T, Brilliantov
N, eds. (Springer, NY).

\noindent Goldhirsch I (2003) Rapid granular flows. In \emph{Annual Review
of Fluid Mechanics,} Vol 35, pp 267-293.

\noindent Brilliantov N, P\"{o}schel T (2004) \emph{Kinetic Theory of
Granular Gases}, (Oxford, New York).

\noindent (2004) \emph{The Physics of Granular Media}, Hinrichsen H, Wolf D,
eds., (Wiley-VCH, Berlin).

\noindent(2004) \emph{Unifying Concepts in Granular Media and Glasses},
Coniglio A, Fierro A, Herrmann H, Nicodemi M, eds. (Elsevier, Amsterdam).

\noindent P\"{o}schel T, Schwager T (2005) \emph{Computational Granular
Dynamics : Models and Algorithms}, (Springer, NY).

\noindent Dufty JW (2007) Nonequilibrium statistical mechanics and
hydrodynamics for a granular fluid. Six lectures at the Second Warsaw School
on Statistical Physics, Kazimierz, Poland, arXiv:0707.3714

\appendix\vspace{0.3in}

\noindent\textbf{Appendix - Gradient expansion}

\label{appA}In this Appendix the Liouville equation in the form (\ref{6.5})
is written to first order in the gradients and solved. Also the invariants
of the associated dynamics are identified.

Consider first the right side of (\ref{6.5}) which can be written
equivalently as%
\begin{eqnarray}
&&\int d\mathbf{r}^{\prime }\frac{\delta \rho _{0\ell }}{\delta \left\langle
a_{\alpha }\left( \mathbf{r}^{\prime }\right) ;t\right\rangle }\left\{
\nabla \cdot \left\langle \mathbf{b}_{\alpha }\left( \mathbf{r}\right)
;t\right\rangle +\delta _{\alpha 2}\left\langle w\left( \mathbf{r}\right)
;t\right\rangle \right\} -\overline{L}\rho _{0\ell }  \notag \\
&=&-\int d\mathbf{r}^{\prime }\frac{\delta \rho _{0\ell }}{\delta
\left\langle a_{\alpha }\left( \mathbf{r}^{\prime }\right) ;t\right\rangle }%
\left\langle La_{\alpha }\left( \mathbf{r}^{\prime }\right) ;t\right\rangle -%
\overline{L}\rho _{0\ell }  \notag \\
&=&\int d\mathbf{r}^{\prime }\frac{\delta \rho _{0\ell }}{\delta
\left\langle a_{\alpha }\left( \mathbf{r}^{\prime }\right) ;t\right\rangle }%
\int d\Gamma a_{\alpha }\left( \mathbf{r}\right) \overline{L}\left( \rho
_{0\ell }+\Delta \right) -\overline{L}\rho _{0\ell }  \label{a.1}
\end{eqnarray}%
The first equality follows from (\ref{4.2}) and (\ref{5.1}). The first two
terms are determined by the local HCS which can be expanded to first order
in the gradients%
\begin{eqnarray}
\rho _{0\ell } &=&\rho _{0}\left( \left\langle a_{\alpha }\left( \mathbf{r}%
\right) ;t\right\rangle \right) +\int d\mathbf{r}^{\prime }\left( \frac{%
\delta \rho _{0\ell }}{\delta \left\langle a_{\alpha }\left( \mathbf{r}%
^{\prime }\right) ;t\right\rangle }\right) _{\delta \left\langle a_{\alpha
};t\right\rangle =0}\left( \left\langle a_{\alpha }\left( \mathbf{r}^{\prime
}\right) ;t\right\rangle -\left\langle a_{\alpha }\left( \mathbf{r}\right)
;t\right\rangle \right) +\cdot \cdot  \notag \\
&=&\rho _{0}\left( \left\langle a_{\alpha }\left( \mathbf{r}\right)
;t\right\rangle \right) +\mathbf{m}_{\beta }\left( \mathbf{r},\left\langle
a_{\alpha }\left( \mathbf{r}\right) ;t\right\rangle \right) \cdot \nabla
\left\langle a_{\beta }\left( \mathbf{r}\right) ;t\right\rangle +\cdot \cdot
\label{a.2}
\end{eqnarray}%
The functional derivatives are%
\begin{eqnarray}
\left( \frac{\delta \rho _{0\ell }}{\delta \left\langle a_{\alpha }\left(
\mathbf{r}^{\prime }\right) ;t\right\rangle }\right) _{\delta \left\langle
a_{\alpha };t\right\rangle =0} &=&\delta \left( \mathbf{r}^{\prime }\mathbf{%
-r}\right) \left( \frac{\partial \rho _{0}\left( \left\langle a_{\alpha
}\left( \mathbf{r}\right) ;t\right\rangle \right) }{\partial \left\langle
a_{\alpha }\left( \mathbf{r}\right) ;t\right\rangle }+\frac{\partial \mathbf{%
m}_{\beta }\left( \mathbf{r},\left\langle a_{\alpha }\left( \mathbf{r}%
\right) ;t\right\rangle \right) }{\partial \left\langle a_{\alpha }\left(
\mathbf{r}\right) ;t\right\rangle }\cdot \nabla \left\langle a_{\beta
}\left( \mathbf{r}\right) ;t\right\rangle \right)  \notag \\
&&+\mathbf{m}_{\alpha }\left( \mathbf{r},\left\langle a_{\alpha }\left(
\mathbf{r}\right) ;t\right\rangle \right) \cdot \nabla \delta \left( \mathbf{%
r}^{\prime }\mathbf{-r}\right) +\cdot \cdot  \label{a.3}
\end{eqnarray}%
Here,%
\begin{equation}
\mathbf{m}_{\beta }\left( \mathbf{r},\left\langle a_{\alpha }\left( \mathbf{r%
}\right) ;t\right\rangle \right) \equiv \int d\mathbf{r}^{\prime }\left(
\frac{\delta \rho _{0\ell }}{\delta \left\langle a_{\beta }\left( \mathbf{r}%
^{\prime }\right) ;t\right\rangle }\right) _{\delta \left\langle a_{\alpha
};t\right\rangle =0}\left( \mathbf{r}^{\prime }\mathbf{-r}\right) ,
\label{a.4}
\end{equation}%
and $\rho _{0}\left( \left\langle a_{\alpha }\left( \mathbf{r}\right)
;t\right\rangle \right) $ is the actual HCS with its global density, energy,
and momentum evaluated at the common values $\left\langle a_{\alpha }\left(
\mathbf{r}\right) ;t\right\rangle $. It follows from (\ref{6.2}) that the
averages of $a_{\alpha }\left( \mathbf{r}\right) $ for $\rho ,\rho _{0\ell
}, $ and $\rho _{0}$ are all the same. This in turn gives
\begin{equation}
\int d\Gamma a_{\alpha }\left( \mathbf{r}\right) \mathbf{m}_{\beta }=0=\int
d\Gamma a_{\alpha }\left( \mathbf{r}\right) \Delta .  \label{a.5}
\end{equation}

With these results and the fact that $\Delta $ is of first order in the
gradients, (\ref{a.1}) to first order in the gradients becomes%
\begin{eqnarray}
&&\int d\mathbf{r}^{\prime }\frac{\delta \rho _{0\ell }}{\delta \left\langle
a_{\alpha }\left( \mathbf{r}^{\prime }\right) ;t\right\rangle }\left\{
\nabla \cdot \left\langle \mathbf{b}_{\alpha }\left( \mathbf{r}\right)
;t\right\rangle +\delta _{\alpha 2}\left\langle w\left( \mathbf{r}\right)
;t\right\rangle \right\} -\overline{L}\rho _{0\ell }  \notag \\
&\rightarrow &\overline{\mathcal{L}}\rho _{0}-\left( 1-\mathcal{P}\right)
\left( I\overline{\mathcal{L}}+K^{T}\right) _{\alpha \beta }\mathbf{m}%
_{\beta }\cdot \nabla \left\langle a_{\alpha };t\right\rangle +\mathcal{P}%
\overline{\mathcal{L}}\Delta  \label{a.6}
\end{eqnarray}%
The matrix $K^{T}$ is the transpose of $K$%
\begin{equation}
K_{\alpha \beta }=\delta _{\alpha 2}\frac{\partial \omega (\left\langle
a_{\alpha }\left( \mathbf{r}\right) ;t\right\rangle )}{\partial \left\langle
a_{\beta }\left( \mathbf{r}\right) ;t\right\rangle },  \label{a.7}
\end{equation}%
and $I$ is the unit matrix. The generator $\overline{\mathcal{L}}$ is the
same as that of (\ref{3.19}) with $\omega \rightarrow \omega
_{0}(\left\langle a_{\alpha }\left( \mathbf{r}\right) ;t\right\rangle )$ for
the HCS evaluated at the common values $\left\langle a_{\alpha }\left(
\mathbf{r}\right) ;t\right\rangle $%
\begin{equation}
\overline{\mathcal{L}}=-\omega _{0}(\left\langle a_{\alpha }\left( \mathbf{r}%
\right) ;t\right\rangle )\partial _{\left\langle e\left( \mathbf{r}\right)
;t\right\rangle }+\overline{L}.  \label{a.8}
\end{equation}%
Finally, $\mathcal{P}$ is the projection operator%
\begin{equation}
\mathcal{P}X=\frac{\partial \rho _{0}}{\partial \left\langle a_{\alpha
}\left( \mathbf{r}\right) ;t\right\rangle }\int d\Gamma a_{\alpha }\left(
\mathbf{r}\right) X.  \label{a.9}
\end{equation}%
The first term of (\ref{a.6}) vanishes by definition of the HCS, $\rho _{0}$%
, confirming that the right side of the Liouville equation (\ref{6.5}) is of
first order in the gradients.

At this point, the Liouville equation (\ref{6.5}) becomes
\begin{equation}
\partial _{t}\Delta -\int d\mathbf{r}^{\prime }\frac{\delta \Delta }{\delta
\left\langle a_{2}\left( \mathbf{r}^{\prime }\right) ;t\right\rangle }\omega
_{0}(\left\langle a_{\alpha }\left( \mathbf{r}^{\prime }\right)
;t\right\rangle )+\mathcal{P}\overline{L}\Delta =\left( 1-\mathcal{P}\right)
\boldsymbol{\Upsilon }_{\alpha }^{\prime }\cdot \nabla \left\langle
a_{\alpha };t\right\rangle ,  \label{a.10}
\end{equation}%
\begin{equation}
\boldsymbol{\Upsilon }_{\alpha }^{\prime }\equiv -\left( I\overline{\mathcal{%
L}}+K^{T}\right) _{\alpha \beta }\mathbf{m}_{\beta }  \label{a.11}
\end{equation}%
This equation is still exact up through contributions of first order in the
gradients. It has solutions of the form%
\begin{equation}
\Delta \left( \Gamma ,t\mid \left\langle a_{\alpha }\left( \mathbf{r}\right)
;t\right\rangle \right) =\mathbf{G}_{\nu }\left( \Gamma ,t,\left\langle
a_{\alpha }\left( \mathbf{r}\right) ;t\right\rangle \right) \cdot
\boldsymbol{\nabla }\left\langle a_{\nu }\left( \mathbf{r}\right)
;t\right\rangle ,  \label{a.12}
\end{equation}%
Substitution into (\ref{a.10}) gives the corresponding equation for $\mathbf{%
G}_{\nu }$%
\begin{equation}
\partial _{t}\mathbf{G}_{\nu }+\left( 1-\mathcal{P}\right) \left( I\overline{%
\mathcal{L}}+K^{T}\right) _{\nu \beta }\mathbf{G}_{\beta }=\left( 1-\mathcal{%
P}\right) \boldsymbol{\Upsilon }_{\nu }^{\prime },  \label{a.13}
\end{equation}%
with the solution%
\begin{equation}
\mathbf{G}_{\nu }\left( \Gamma ,t,\left\langle a_{\alpha }\left( \mathbf{r}%
\right) ;t\right\rangle \right) =\int_{0}^{t}dt^{\prime }\left( e^{-\left( 1-%
\mathcal{P}\right) \left( I\overline{\mathcal{L}}+K^{T}\right) t^{\prime
}}\right) _{\nu \beta }\left( 1-\mathcal{P}\right) \boldsymbol{\Upsilon }%
_{\beta }^{\prime }.  \label{a.14}
\end{equation}%
It is possible to add to (\ref{a.10}) an arbitrary solution to the
homogeneous equation corresponding to (\ref{a.13}). As described in the
text, this represents the dynamics of the first stage of rapid velocity
relaxation to the local HCS. The interest here is in the second stage where
possible formation of a normal solution occurs. Hence, it is simpler to
choose that stage for initial conditions (initial local HCS).

Define the derivatives of the HCS by%
\begin{equation}
\Psi _{\beta }\left( \Gamma ,\left\langle a_{\alpha }\left( \mathbf{r}%
\right) ;t\right\rangle \right) \equiv \frac{\partial \rho _{0}}{\partial
\left\langle a_{\beta }\left( \mathbf{r}\right) ;t\right\rangle }.
\label{a.15}
\end{equation}%
Then differentiate the equation for $\rho _{0}$%
\begin{equation}
\frac{\partial }{\partial \left\langle a_{\beta }\left( \mathbf{r}\right)
;t\right\rangle }\overline{\mathcal{L}}\rho _{0}=0,  \label{a.16}
\end{equation}%
to get%
\begin{equation}
\left( I\overline{\mathcal{L}}_{T}+K^{T}\right) _{\nu \beta }\Psi _{\beta
}=0.  \label{a.17}
\end{equation}%
Since $\left( I\overline{\mathcal{L}}_{T}+K^{T}\right) $ is the generator
for the dynamics in (\ref{a.14}) this shows that $\Psi _{\beta }$ are the
invariants of that dynamics.

The projection operator $\mathcal{P}$ in (\ref{a.18}) acts only on phase
functions with translational invariance. In that case (\ref{a.9}) simplifies
to%
\begin{equation}
\mathcal{P}X=\Psi _{\beta }\int d\Gamma A_{\beta }X,\hspace{0.3in}A_{\beta
}=V^{-1}\int d\mathbf{r}a_{\alpha }\left( \mathbf{r}\right) .  \label{a.18}
\end{equation}%
The first equality of (\ref{a.5}) becomes $\mathcal{P}\mathbf{m}_{\beta }=0$%
. This in turn gives
\begin{equation}
\mathbf{m}_{\beta }=\left( 1-\mathcal{P}\right) \mathbf{m}_{\beta }=\left( 1-%
\mathcal{P}\right) \mathbf{M}_{\beta },\hspace{0.3in}\mathbf{M}_{\beta
}\equiv \int d\mathbf{r}^{\prime }\left( \frac{\delta \rho _{0\ell }}{\delta
\left\langle a_{\alpha }\left( \mathbf{r}^{\prime }\right) ;t\right\rangle }%
\right) _{\delta \left\langle a_{\alpha };t\right\rangle =0}\mathbf{r}%
^{\prime }.  \label{a.18a}
\end{equation}%
Then $\left( 1-\mathcal{P}\right) \boldsymbol{\Upsilon }_{\alpha }^{\prime }$
simplifies to%
\begin{equation}
\left( 1-\mathcal{P}\right) \boldsymbol{\Upsilon }_{\alpha }^{\prime
}=-\left( 1-\mathcal{P}\right) \left( I\overline{\mathcal{L}}+K^{T}\right)
_{\alpha \beta }\left( 1-\mathcal{P}\right) \mathbf{M}_{\beta }\equiv \left(
1-\mathcal{P}\right) \boldsymbol{\Upsilon }_{\alpha }  \label{a.19}
\end{equation}%
with%
\begin{equation}
\boldsymbol{\Upsilon }_{\alpha }=-\left( I\overline{\mathcal{L}}%
+K^{T}\right) _{\alpha \beta }\mathbf{M}_{\beta }.  \label{a.20}
\end{equation}%
Use has been made of the identity%
\begin{equation}
\left( 1-\mathcal{P}\right) \left( I\overline{\mathcal{L}}+K^{T}\right)
\mathcal{P}=0.  \label{a.21}
\end{equation}%
This same identity leads to a simplification of the dynamics in (\ref{a.14})%
\begin{equation}
e^{-\left( 1-\mathcal{P}\right) \left( I\overline{\mathcal{L}}+K^{T}\right)
t^{\prime }}\left( 1-\mathcal{P}\right) =\left( 1-\mathcal{P}\right)
e^{-\left( I\overline{\mathcal{L}}+K^{T}\right) t^{\prime }}\left( 1-%
\mathcal{P}\right) .  \label{a.22}
\end{equation}

In summary, the solution to the Liouville equation to first order in the
gradients is%
\begin{eqnarray}
\rho \left( \Gamma ,t\mid \left\langle a_{\alpha };t\right\rangle \right)
&=&\rho _{0}\left( \Gamma ,\left\langle a_{\alpha }(\mathbf{r)}%
;t\right\rangle \right) \mathbf{+}\left( 1-\mathcal{P}\right) \left( \mathbf{%
M}_{\beta }\left( \Gamma ,\left\langle a_{\alpha }(\mathbf{r)}%
;t\right\rangle \right) \right.  \notag \\
&&\left. +\int_{0}^{t}dt^{\prime }\left( e^{-\left( I\overline{\mathcal{L}}%
+K^{T}\right) t^{\prime }}\right) _{\nu \beta }\left( 1-\mathcal{P}\right)
\boldsymbol{\Upsilon }_{\beta }\left( \Gamma ,\left\langle a_{\alpha }(%
\mathbf{r)};t\right\rangle \right) \right) \cdot \nabla \left\langle a_{\nu
}\left( \mathbf{r}\right) ;t\right\rangle  \notag \\
&&  \label{a.23}
\end{eqnarray}


\begin{thebibliography}{99}
\bibitem{Navier-Stokes} Here the Navier-Stokes approximation is defined by
calculating the cooling rate, energy flux, and momentum flux to first order
in the gradients. However, the fluxes occur under a gradient in the
macroscopic balance equations while the cooling rate does not. Hence the
equations themselves do not have all terms to second order in the gradients
(i.e., the additional terms of second order contributing to the cooling
rate).
\end{thebibliography}
\end{document}